\documentstyle[aps,prb,preprint,tighten]{revtex}

\newcommand{\nn}[1]{{\langle #1 \rangle}}
\newcommand{\qave}[2]{\left\langle #1 \left| #2 \right| #1 \right\rangle}

\newcommand{\bra}[1]{\langle #1 |}
\newcommand{\ket}[1]{| #1 \rangle}
\newcommand{\qn}[1]{#1^\dagger #1}

\newcommand{\bk}{{\bf k}}
\newcommand{\bkp}{{{\bf k}'}}
\newcommand{\bq}{{\bf q}}
\newcommand{\bS}{{\bf S}}
\newcommand{\bQ}{{\bf Q}}
\newcommand{\spin}{s}
\newcommand{\bspin}{{\bf s}}

\newcommand{\tj}{\mbox{$t$-$J$} }

\begin{document}

\author{A. V. Dotsenko}

\title{Spin polarons in the \tj model in an unconstrained representation}

\address{School of Physics, The University of New South Wales,
         Sydney 2052, Australia\protect\cite{mpi-pks}}

\maketitle

\begin{abstract}
The report
  discusses the slave-fermion representations of the \tj model
  and describes another representation,
  in which fermions and bosons
  are completely commuting
  and in which the properties of fermions
  are directly related to the properties of physical holes.
For a study of the system in the new representation at half-filling,
  interaction of fermions with two magnons
  is treated in mean-field theory.
The obtained effective model,
  in comparison to that of the usual slave-fermion representation,
  has an additional bare hole dispersion
  due to the hole moving
  by using quantum spin fluctuations present
  in the undoped antiferromagnetic ground state.
The single-hole Green's function at half-filling
  is then found numerically
  using the self-consistent Born approximation.
For all studied quantities
  good or excellent agreement with numerical data
  is observed in the entire parameter range,
  noticeably better
  than in the studies with the slave-fermion representation.
Using the same effective model,
  the two-hole problem
  is also studied by solving numerically
  the Bethe-Salpeter equation with noncrossing diagrams.
\end{abstract}

\section{Introduction} \label{sec:intro}

Much, if not most, of the progress in describing theoretically
  the complex physics of high-temperature superconductors
  and strongly correlated electrons
  has been achieved using numerical methods,\cite{dagotto94}
  while analytical methods have been rather approximate
  and not often checked against numerical data.
A simple and transparent analytical or semianalytical model
  can however be very valuable by providing physical insights
  for {\em understanding\/} in addition to describing the system
  and by being easily extendible to other problems.
Currently this
  is the role of
  the so-called self-consistent Born approximation,%
\cite{KLR89,martinez-horsch91,marsiglio-etal91,liu-manousakis}
  which provides a fairly good description of the undoped \tj model.
More accurately,
  in this method
  $(0)$
  the \tj model
  is expressed in terms of holons and bosons\cite{SVR88};
  then $(i)$ the boson part is solved to leading order in $1/S$
     (spin-wave theory);
  and finally $(ii)$ the interacting holon-boson system
  is solved numerically using leading-order diagrams.
(For points $i$ and $ii$, higher-order terms/diagrams
   have been analyzed by
   Liu and  Manousakis.\cite{liu-manousakis})

In this publication I
  report a study of the \tj model in a different scheme.
The Hilbert space of holons and bosons
  is expanded,
  so that they are no longer constrained by each other.
Another important difference
  is that
  the fermionic Green's function in the expanded model
  corresponds directly
  to the Green's function of fermions in the original \tj model.

Numerical results
  are then obtained for the \tj model
  on the square lattice at half-filling.
For the first and simplest stage of analysis
  in the new representation,
  it is suggested to use mean-field treatment
  for zero- and two-magnon terms in the Hamiltonian.
The effective model obtained this way,
  consisting of interacting holes and spin waves,
  is the same as in the usual constrained representation
  with the constraint ignored
  except for the presence of a bare hole dispersion
  due to the hole moving
  by eating away spin fluctuations present in the ground state.
Solving the equations of motion for one hole
  in the self-consistent Born approximation,
  I find a fairly good agreement with numerical results
  for all analyzed quantities,
  such as the bandwidth and the band structure,
  and for all parameter regimes.
The two-hole problem
  is also studied using the leading-order, noncrossing diagrams,
  demonstrating again viability of the method
  albeit much less convincingly.

In the following Sec.~\ref{sec:transform},
  I describe the discussed formulations of the \tj model.
Then, in Sec.~\ref{sec:results}
  the obtained results for single- and two-hole problems
  are compared against the available numerical data
  and against the results obtained
  in the usual slave-fermion representation.
The report ends with a summary (Sec.~\ref{sec:summary}).

\section{Analytical transformations of the \tj model} \label{sec:transform}

The familiar \tj Hamiltonian is
\begin{equation}
\label{eq:Htj}
  H_{tJ} = - t \sum_{\nn{ij}\sigma} (c_{i\sigma}^\dagger  c_{j\sigma}
				      + \text{H.c.})
	   + J \sum_\nn{ij} \left( \bS_i \cdot \bS_j - \case{1}{4} n_i n_j 
			     \right),
\end{equation}
  where the notation is standard
  with the addition that, both here and throughout the paper,
  $n$, $i$, and $j$ refer to
  any, spin-up, and spin-down sublattice sites, respectively
  ($n$ as a subindex
   is not to be confused with the electron number operators $n_n$).
The square lattice is implied,
  although much of the discussion is independent of this.
Recently, an extra $t'$ term describing next-nearest-neighbor hopping
  is usually added to the model.\cite{andersen-etal95}
This study however,
  being mainly for demonstrative and comparative purposes,
  is restricted to the original \tj model Eq.~(\ref{eq:Htj}).

The Hamiltonian Eq.~(\ref{eq:Htj})  
  is written without the commonly included
  electron projection operators
  on the understanding that, instead,
  the Hilbert space
  is restricted to no double electron occupancy
  (see Fig.~\ref{fig:hilbert}).
It also goes without saying that the ``electrons'' of the \tj model
  are in fact Zhang-Rice singlets.\cite{zhang-rice88}

\subsection{The holon-{\em or}-spin representation of the \tj model}

To disentangle the Hamiltonian,
  the main degrees of freedom
  must be identified
  and suitable operators introduced.
Intuitively, spin fluctuations and mobile holes
  are the main underlying objects.
It was, to the best of the author's knowledge,
  Schmitt-Rink, Varma, and Ruckenstein\cite{SVR88}
  who first proposed to represent the system as a combination
  of a spinless-fermion (holon) field and a boson field.
In this slave-fermion representation,
  the electron operators
  are written as
  $c_{n\sigma} = f_n^\dagger b_{n\sigma}$
  and the Hilbert space of holons $f_n$ and bosons $b_{n\sigma}$
  is constrained by
  $\qn{f_n} + \qn{b_{n\uparrow}} + \qn{b_{n\downarrow}} = 1$.

We can think of bosons $b_{n\sigma}$
  as the Schwinger bosons of some spin field $\bspin_n$,
  which in this case
  corresponds directly to the physical spin, $\bS_n = \bspin_n$.
This spin field $\bspin_n$
  can also be of course represented by another type of bosons,
  such as the bosons $a_n$
  of the Holstein-Primakoff or Dyson-Maleev transformation.
The entire system
  is then represented in terms of operators $f_n$ and $a_n$
  (in the normally implied case of antiferromagnetic order
   the bosonic operators $a_n$ are different on the two sublattices).
The Hilbert space
  is constrained in this case by
  $\qn{f_n} + \qn{a_n} < 2$.

Finally, the spin field $\bspin_n$
  can be present directly rather than via bosons.
This corresponds to what
  was done in Ref.~\onlinecite{richard-yushankhai93}
  and is described here in some detail since
  it is closer to the model presented later.
The \tj model Hilbert space Fig.~\ref{fig:hilbert}
  is mapped onto the one in Fig.~\ref{fig:holonHilbert}.
At any site $n$, there can be a holon $f_n$
  and there is {\em always\/} a spin $\bspin_n$.
A ``normal,'' that is containing one electron, site
  is thought of as having no holon,
  while the spin field on such site
  is identified with the physical electron spin,
  $\bspin_n = \bS_n$ for $\qn{f_n} = 0$.
An empty site
  is considered to have one holon.
The spin $\bspin_n$ on such a site
  is a dummy, or a ghost---it is unphysical.
We choose to make it
  {\em up\/} if the site is on the spin-up sublattice
  and {\em down\/} otherwise.
This choice
  is arbitrary
  and is motivated by the Ising limit.

The relations for operators
  are as follows.
The physical spin
  is $\bS_n = (1 - \qn{f_n}) \, \bspin_n$.
The electron operators
  are
  $c_{i\uparrow} = (\case{1}{2} + s_{i}^{z}) f_i^\dagger$
  and $c_{i\downarrow} = s_{i}^{+} f_i^\dagger$ for the spin-up sublattice
  and
  $c_{j\uparrow} = s_{j}^{-} f_j^\dagger$
  and $c_{j\downarrow} = (\case{1}{2} - s_{j}^{z}) f_j^\dagger$
  for the spin-up sublattice.
The \tj Hamiltonian
  is then in this representation (ignoring spin projection operators)
\begin{equation}
  H_{tJ} = t \sum_{\nn{ij}} f_i^\dagger f_j
			    (s_{j}^{+} + s_{i}^{+} + \mbox{H.c.})
	   + J \sum_\nn{ij} (1 - \qn{f_i})
				\left( \bspin_i \cdot \bspin_j
					- \case{1}{4}
				\right)
			      (1 - \qn{f_j}).
\end{equation}
Again, we have an inhomogeneous Hilbert space,
  in which a spin deviation and a holon
  cannot coexist on a site.
This transformation
  can be generalised to
  $c_{n\sigma} = f_n^\dagger \hat{U}_s$,
  where $\hat{U}_s$
  is some spin transformation with appropriate properties.

The representations described so far
  constitute big progress relative to the ``raw'' \tj model.
In such analytical representations,
  by use of the half-filled state as the background,
  most correlations present in the system
  are already taken into account.
Due to fermion statistics,
  it is automatically guaranteed
  that not more than one holon can be on a site.
However there are also problems.

(1) The first one is the constraint,
  which any not overly complicated wave functions
  can hopefully satisfy only on average.
Of course, we may ``get rid'' of the constraint by
  introducing into the Hamiltonian projection operators,
  but then the Hamiltonian becomes complicated.
It will then have some artificial
  complicated terms of purely ``kinematic'' origin.

(2) The second disadvantage
  is that the holon operators
  in these representations
  only approximately correspond to the original electron operators.
Indeed, $c_{n\uparrow}$
  is defined differently on different sublattices,
  so that $c_{\bk\uparrow}$
  is not the same as $f_\bk^\dagger$
  but rather is a combination of $f_\bk^\dagger$
  and $f_{\bk-\bq}^\dagger s_\bq^{-}$.
It is then necessary to do additional calculations
  to find the Green's function of operators $c_n$
  (see Ref.~\onlinecite{sushkov-etal97}
   and also the Appendix of Ref.~\onlinecite{martinez-horsch91}).

Physically, both points (1) and (2)
  are related to ground-state quantum spin fluctuations
  and dissappear completely for a classical N{\'e}el background.
In fact even for the quantum N{\'e}el background
  as long as {\em linear\/} spin-wave theory
  is used,
  the constraint
  can be formally ignored.
Both problem however
  get quickly more dramatic as antiferromagnetic order is weakened,
  which is known to happen in copper oxides.

There is also
  a problem with analyzing the spin field by itself.
Spin-wave theory
  is certainly only approximate and has its limitations.
However that is
  a general problem of condensed-matter physics
  rather than a problem with any particular representation of the \tj model.

\subsection{The unconstrained representation of the \tj model}

It is possible
  to study the \tj model
  in a substantially different representation.
The central idea
  is to relax the constraint, so that
  when there is a hole on a site
  the (ghost) spin on that site
  can have {\em any\/} value.

This is a big step.
The new Hilbert space
  has four states per site
  instead of three in the original model.
Instead of being able to have
  strictly a holon {\em or\/} a boson,
  we can have a holon {\em and/or\/} boson.
Thus we have essentially a different, a bigger model.
No one-to-one transformation is taking place.
(It has in fact been claimed recently\cite{kochetov-yarunin97}
  using symmetry arguments
  that it is impossible to transform
  the \tj model to a model of commuting fermions and spins.)
We may still
  call the new model a ``representation'' of the \tj model
  in the sense of group theory (a reducible representation).
Two of the states in the new model,
  $\ket{0_f, \uparrow}$ and $\ket{0_f, \downarrow}$,
  correspond to one state in the original model, $\ket{\cdot}$.
The correspondence for spin operators
  is as simple as before,
  $\bS_n \rightarrow (1 - \qn{f_n}) \, \bspin_n$.
It is also clear
  that a hole remains a hole.
For single-fermion operators
  there is however no simple relation.

We are free
  to define
  the dynamics of the ghost spin in new model---one
  can think of it as choosing the gauge.
(The dynamics of the nonghost spin
  should, of course, be the same as in the original model.)
For example,
  we can choose the hopping part of the Hamiltonian
  to have the following form,
\begin{equation}
  H_t = t \sum_{\nn{nn'}\sigma \sigma'}
        \left( \ket{1_f \sigma'}_{n'} \ket{0_f \sigma}_n
		\bra{1_f \sigma'}_n    \bra{0_f \sigma}_{n'}
		+ \mbox{H.c.}
	\right).
\end{equation}
Here $\sigma$
  is a nonghost spin,
  and so it should just hop without flipping.
We have chosen
  that the ghost spin, $\sigma'$, also simply hops.
The ghost spin
  is thus permanently attached to the hole.
Note that it is important here
  to consider hopping
  as a single elementary process,
  not as a combination
  of two processes of destruction and creation of the hole.

The complete Hamiltonian,
  when written in terms of spin operators,
  is then
\begin{equation}
\label{eq:HWangRice}
  H_{f\bspin} = 2 t \sum_\nn{nn'} f_n^\dagger f_{n'}
                            \left(\case{1}{4} + \bspin_n \cdot \bspin_{n'}
                            \right)
          + J \sum_\nn{nn'} (1 - \qn{f_n})
			     \left(\bspin_n \cdot \bspin_{n'} - \case{1}{4}
			     \right)
			     (1 - \qn{f_{n'}}).
\end{equation}
The subindex $f\bspin$ at $H$
  emphasises that this Hamiltonian
  is not the \tj Hamiltonian.
$H_{f\bspin}$ is different from $H_{tJ}$,
  it acts in a different Hilbert space,
  in a Hilbert space that it is
  not even isomorphic
  to the Hilbert space of the \tj Hamiltonian.
Nevertheless $H_{fs}$
  has been constructed in such a way
  that the properties in the \tj model
  can be simply derived
  from the properties in the ``$f$-$\bspin$'' model.

The Hamiltonian Eq.~(\ref{eq:HWangRice})
  has been derived by Wang and Rice.\cite{wang-rice94}
However I believe
  that their method of derivation
  is doubtful
  and their interpretation
  is different in that they claim to have
  an exact expression of the \tj model
  ({\em i.e.\/} a one-to-one transformation).
In any case,
  there have been no calculation in this representation.

The spin-fermion interaction in Eq.~(\ref{eq:HWangRice})
  is of a very natural form for spinless fermions.
The fermion
  propagates either by emitting and absorbing two magnons
  or directly by using fluctuations present in the ground state.
The opposite cases
  of completely antiferromagnetic and ferromagnetic order
  are naturally covered.

Although quite simple and very symmetric,
  this model
  is relatively difficult to study
  because it leads to two-magnon processes.
Below,
  the model with ghost-spin dynamics defined in a different way
  will be analysed.
The dynamics of the ghost spin at hopping processes
  is defined by the diagrams in Fig.~\ref{fig:processes}.
The logic behind these relatively complicated rules
  is clarified in Appendix~\ref{sec:appendix}
  where an Ising-type situation is considered.
The intention
  is to have two-magnon-type interaction for
  the relatively-infrequent interaction with ground-state fluctuations
  but single-magnon for creating fluctuations.

Now it is time to see if
  the problems (1) and (2) of the slave-fermion representation
  have been answered.

(1)
It is important
  that there are
  now no boson-fermion constraints,
  the two fields
  are interacting dynamically
  but are {\em kinematically\/} completely free
  from each other.
Their might be
  constraints on the bosons when the spin field $\bspin_n$
  is expressed through them.
However this is a constraint, {\em e.g.\/}, from spin-wave theory
  and it concerns bosons alone.
The new model
  does not solve the problem of magnetic order as such,
  but it does separate it from the total problem.

(2)
To prove that the properties of fermions in this model
  are equal
  to the properties of holes in the \tj model,
  consider the following argument.
Let us set {\em at photoemission processes\/}
  the following correspondence:
  $c_{n\downarrow} \rightarrow f_n^\dagger (\case{1}{2} + s_n^z)$
  and
  $c_{n\uparrow} \rightarrow f_n^\dagger (\case{1}{2} - s_n^z)$
  and likewise for Hermitian conjugates.
That is to say that
  creation of a hole
  is represented as creation of a holon
  and, importantly, the spin on the site
  remains unchanged
  (we may say
  that it turns from a nonghost one to a ghost one).
At subsequent hoppings the ghost spin
  changes according
  to the rules outlined above;
  the physics should be independent of this.
Then we have
  $c_{\bk\uparrow} + c_{\bk\downarrow} \rightarrow f_\bk^\dagger$
  and it is easy to see that there will be a direct relation for
  Green's functions.
The holon
  is now in {\em some\/} sense
  a symmetric mixture of spin-up and spin-down holes.

It should be also possible to introduce spin-carrying holes.
The Hilbert space would be the direct product
  of the space of $f_{n\uparrow}$ and $f_{n\downarrow}$
  (coexistence forbidden)
  and of spin $\bspin_n$.
However it is unclear if such a version would give any advantages.

\subsection{Deriving an effective model}

At this point the Hamiltonian
  has only become more complicated.
The important point however
  is that it can now be simplified significantly
  without loosing much accuracy.
Fortunately, having numerical data
  allows to quickly test if this is so,
  at least in simple situations.

The guidance for the simplifications
  is that the spin fluctuations in the ground state
  are present in small numbers.
The well-known\cite{manousakis91}
  characteristics of the N\'eel ground state
  are
  the staggered magnetization,
\begin{math}
  m^\dagger  = | \qave{\psi_{\text{N\'eel}}}{\spin_n^z} |
	 \approx 0.305,
\end{math}
  and the nearest-neighbor spin correlator,
\begin{math}
  e_{\text{AF}}
	= \left.\qave{\psi_{\text{N\'eel}}}{\bspin_i \cdot \bspin_j }
	  \right|_{\nn{ij}}
	\approx -0.33.
\end{math}

The first two processes in Fig.~\ref{fig:processes}
  describe the hole propagating by emitting and absorbing spin waves.
These processes are the same as known for the slave-fermion scheme and
   I treat them to leading order in $1/S$.
The last two processes of Fig.~\ref{fig:processes}
  describe the hole moving ``friction-free,''
  by using the liquid component of the N\'{e}el state.
In these terms I replace spin combinations by
  their expectation values
  ({\em i.e.\/} do a mean-field/Hartree-Fock procedure
   where the inverse influence of the holes on the spin field is ignored).
A relationship required here is
\begin{displaymath}
  \qave{\psi_{\text{N\'{e}el}}} {(\case{1}{2} + \spin_i^z)
				 (\case{1}{2} + \spin_j^z)
				  + \spin_i^+ \spin_j^- }
   =  \case{1}{4} + e_{\text{AF}}.
\end{displaymath}

In the following, the following standard notation is used,
   $z = 4$ is the coordination number,
   $N$ is the number of sites on the lattice,
   $\bQ = (\pi, \pi)$,
   $\gamma_\bk = \case{1}{2} (\cos k_x + \cos k_y)$,
   and $\nu_\bq = \left( 1 - \gamma_\bq^2 \right) ^{1/2}$.

With the described simplifications and after Fourier transformations,
\begin{mathletters}
  the effective Hamiltonian is
\label{eq:Hfinal}
\begin{eqnarray}
  H_{\text{eff}} =
	&& \sum_\bk
                    E^{(0)} (\bk) \, f_{\bk}^\dagger f_\bk
	  + \sum_\bq \omega_\bq \alpha_\bq^\dagger \alpha_\bq
	  + \frac{zt}{\sqrt{N}} \sum_{\bk \bq}
			 M (\bk, \bq)
			(f_{\bk}^\dagger f_{\bk - \bq} \alpha_\bq
			 + \text{H.c.})
							\nonumber\\
	&& \mbox{}
	  + \frac{1}{N} \sum_{\bk \bk' \bq}
			\Gamma_{\text{cont}} (\bq) \,
			f_{\bk' - \bq}^\dagger f_{\bk + \bq}^\dagger
			f_{\bk} f_{\bk'}.
\end{eqnarray}
The first term here,
  as a noticeable difference from the unconstrained case,
  is a bare fermion dispersion
\begin{equation}
  E^{(0)} (\bk)
	= \left( \case{1}{4} - e_{\text{AF}} \right) z J
	   - 2 \left( \case{1}{4} + e_{\text{AF}} \right) z t \gamma_\bk.
\end{equation}
The next two terms, describing spin waves with dispersion
  $\omega_\bq = \case{1}{2} z J \nu_\bq$
  and interaction between holes and spin waves,
  are the same as in the slave-fermion case.
The vertex function can be (conveniently?) written as
\begin{equation}
  M (\bk, \bq)
	= \left[ \case{1}{2} (\nu_\bq^{-1} + 1) \gamma_{\bk - \bq}^2
		 + \case{1}{2} (\nu_\bq^{-1} - 1) \gamma_\bk^2
		 - \nu_\bq^{-1} \gamma_\bq \gamma_\bk \gamma_{\bk - \bq}
	  \right] ^{1/2}.
\end{equation}
The last term, with
\begin{equation}
  \Gamma_{\text{cont}} (\bq)
	= \case{1}{2} \left( e_{\text{AF}} - \case{1}{4} \right)
		z J \gamma_\bq,
\end{equation}
  may be called a ``contact'' interaction
  as it describes instantaneous attraction of holes
  on hearest-neighbor sites
  (easily recognized to be due to the ``broken-bond'' mechanism).
This interaction
  is also present in the slave-fermion case but is usually omitted
  since it is negligible in the usually considered regime of $J/t \ll 1$
  and completely irrelevant in the single-hole problem.
\end{mathletters}
The effective model
  does not cover, obviously,
  the extreme Nagaoka case of $t/J \to \infty$
  and the high-energy physics.

It is trivially seen that the slave-fermion model
  is recovered when switching off the bare hole dispersion.
The linear-in-$t$ hole dispersion in the static limit
  has been known for some time.\cite{dagotto-etal90}
It may be obtained
  directly
  by calculating the amplitudes
\begin{displaymath}
  \qave{\psi_{\text{N\'eel}}}{c_{i\sigma}^\dagger
	\left( t \sum_{\sigma'} c_{j\sigma'}^\dagger c_{i\sigma'} \right)
	c_{j\sigma}},
\end{displaymath}
  and remembering to normalize hole operators.
What is remarkable is that this effect
  can be added ``linearly'' to the spin-wave-emission mechanism.

\section{Comparison of results} \label{sec:results}

\subsection{The single-hole problem}

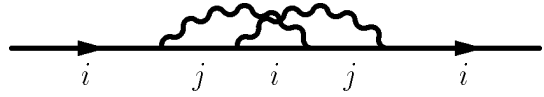
\begin{figure}
  \caption{The leading-order crossing diagram
	for the single-hole Green's function.
	$i$ and $j$ are sublattice indices.}
  \label{fig:crossing}
\end{figure}

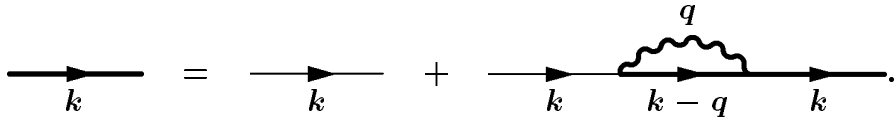
\begin{figure}
  \caption{The Dyson equation for the single-hole Green's function
		in the self-consistent Born approximation.}
  \label{fig:dyson}
\end{figure}

To find the single-hole Green's function
  given the effective Hamiltonian Eqs.~(\ref{eq:Hfinal})
  it appears natural
  to use the same self-consistent Born approximations
  that has been used in the slave-fermion case.
The first crossing diagram
 Fig.~\ref{fig:crossing}
  is prohibited by kinematic reasons
  (it may be seen when the two-sublattice formalism is used
   that the spin wave emitted
   when the hole jumps from sublattice $\{i\}$ to sublattice $\{j\}$
   must be absorbed
   when the hole jumps from sublattice $\{j\}$ back to $\{i\}$).
The Green's function
  is then found by solving
  the following Dyson equation
  (corresponding to the diagram in Fig.~\ref{fig:dyson}),
\begin{equation}
\label{eq:dyson}
  [G (\bk, \omega)]^{-1}
	= [G^{(0)} (\bk, \omega)] ^{-1}
	   - \frac{(zt)^2}{N} \sum_\bq M^2 (\bk, \bq) \,
			  G (\bk - \bq, \omega - \omega_\bq),
\end{equation}
  where $ G^{(0)} (\bk, \omega)
	= [\omega - E^{(0)} (\bk) + i0] ^{-1} $
	is the zeroth order Green's function
  and integration over spin-wave frequencies
  has been carried out
  using the observation that in the single-particle case
  all poles of $G (\bk, \omega)$ are in the lower plane.
(On the antiferromagnetic background,
  we could use the magnetic Brillouin zone,
  but it is more convenient to use the full one,
  it is however not the ``true'' Brillouin zone.)

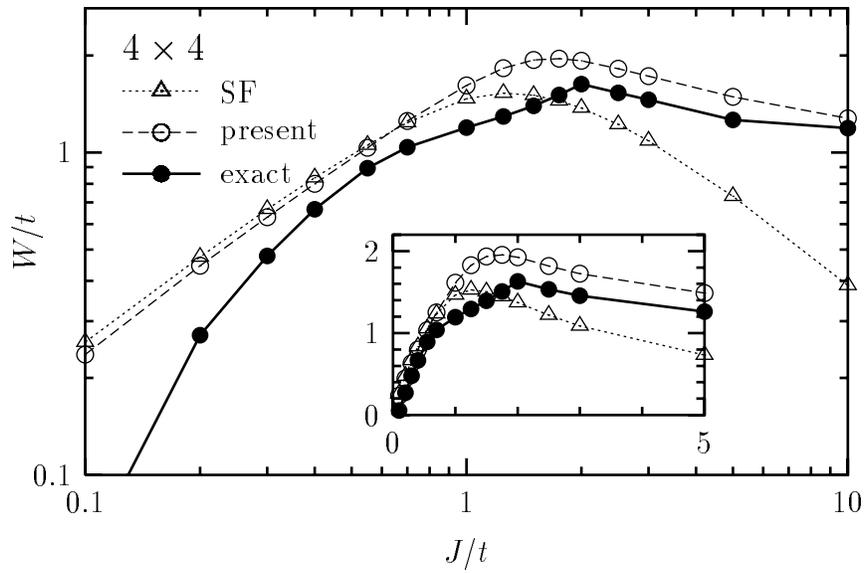
\begin{figure}
  \caption{Quasiparticle bandwidth $W$ on a $4 \times 4$ lattice.}
  \label{fig:bandwidth}
\end{figure}

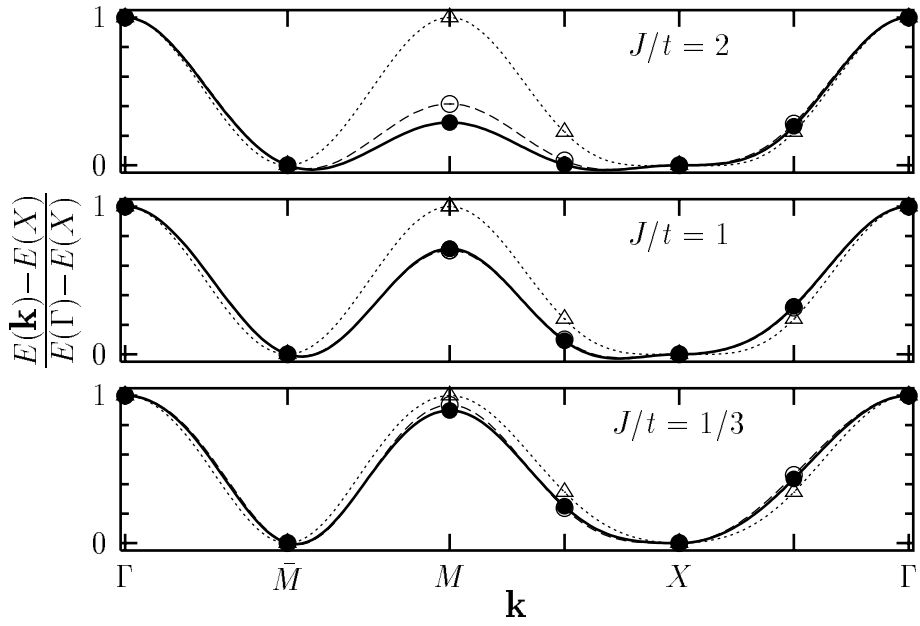
\begin{figure}
  \caption{Quasiparticle band structure on a $4 \times 4$ lattice.
	The lines are trigonometric-function fits to the points plotted.
	The notation is in the previous figure.
	See Fig.~\protect\ref{fig:bandwidth} for the actual scale.
	Note there is no degeneracy
		between {\bf k} and \bk$ + ${\bf Q} observed for holons.
	Except for $J/t = 2$, the results of the present study
		are almost indistinguishable from the exact ones.}
  \label{fig:band}
\end{figure}

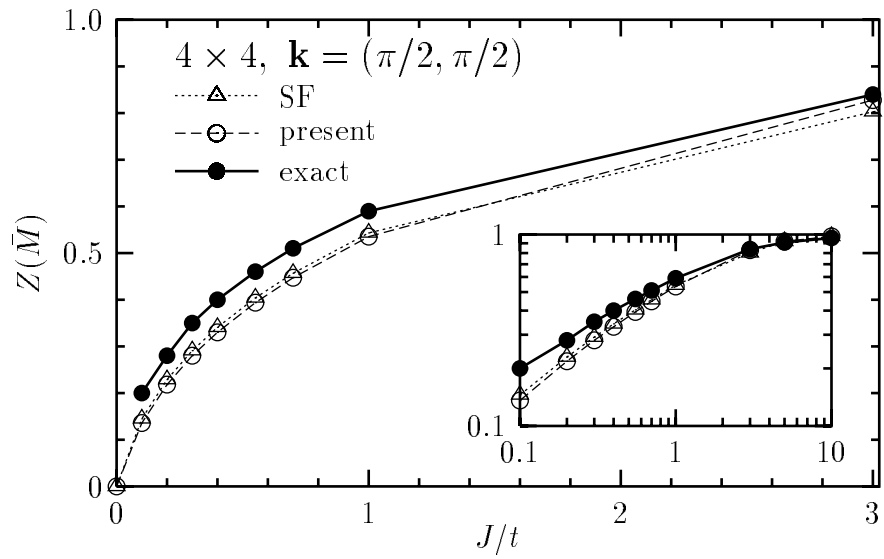
\begin{figure}
  \caption{The quasiparticle residue $Z$
		at the point $\bar{M} = (\pi/2, \pi/2)$
		on a $4 \times 4$ lattice.}
  \label{fig:residue}
\end{figure}

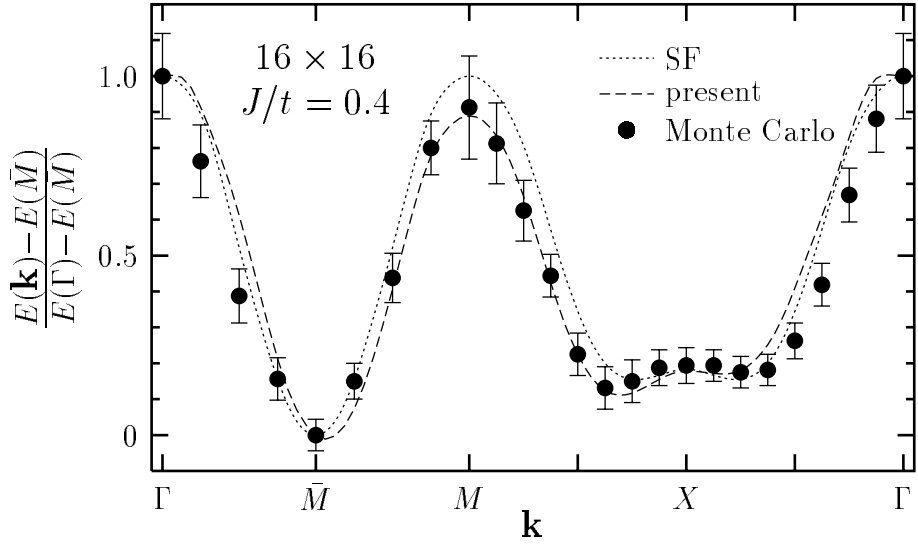
\begin{figure}
  \caption{Quasiparticle band structure at $J/t = 0.4$
		on a $16 \times 16$ lattice.
	The solid line (the result of the present work)
		was smoothed by constructing a trigonometric-function fit
		and it passes through all the points actually calculated.
	The real scales are somewhat different,
		$E(\Gamma) - E(\bar{M}) = 1.18t = 2.95J$
		for the Green's function Monte Carlo
			result\protect\cite{boninsegni93}
		and $E(\Gamma) - E(\bar{M}) = 0.83t = 2.08J$
		for the result of the present study.}
  \label{fig:band16}
\end{figure}

In Figs.\ \ref{fig:bandwidth}, \ref{fig:band}, and \ref{fig:residue},
  the solution of Eq.~(\ref{eq:dyson}) on a $4 \times 4$ lattice
  is compared against exact results
  (I used data
  from Refs.\ \onlinecite{dagotto-etal90}
  and \onlinecite{barnes-etal92}).
The agreement
  can be rated as good to excellent.
Other quantities,
  such as the structure of the spectral function,
  were in good agreement too.
Note also that since no account
  has been taken of small-cluster specifics,
  such as a slightly different magnetic order
  and a relatively large influence of the hole on the spin order
  (very roughly speaking,
   one hole on a $4 \times 4$ lattice
   constitutes a sizeable $6 \%$ doping),
  it may be conjectured
  that the method is even more accurate than
  may be suggested by the given comparison.
That the agreement
  is not an artifact of the highly degenerate $4 \times 4$ lattice
  is proved in Fig.~\ref{fig:band16}, where the dispersion relation
  is compared to that
  obtained\cite{boninsegni93} by the Green's function Monte Carlo method
  on the $16 \times 16$ lattice.
In examining the accuracy
  it should also be remembered
  that physically the low-energy part of the band
  is most important
  (the close agreement for the quasiparticle residue
   at the band bottom
   is thus most encouraging).

Since the computational load
  is quite low,
  it is easy to move on to fairly large lattices.
Finite-size effects per se
  almost disappear starting from the $8 \times 8$ lattice.
The main drawback of small clusters
  seems to be a lack of resolution in $\bk$ space
  (however in the case of dispersion,
   which is a very smooth function,
   it can be mostly overcome
   by using trigonometric-function fits).

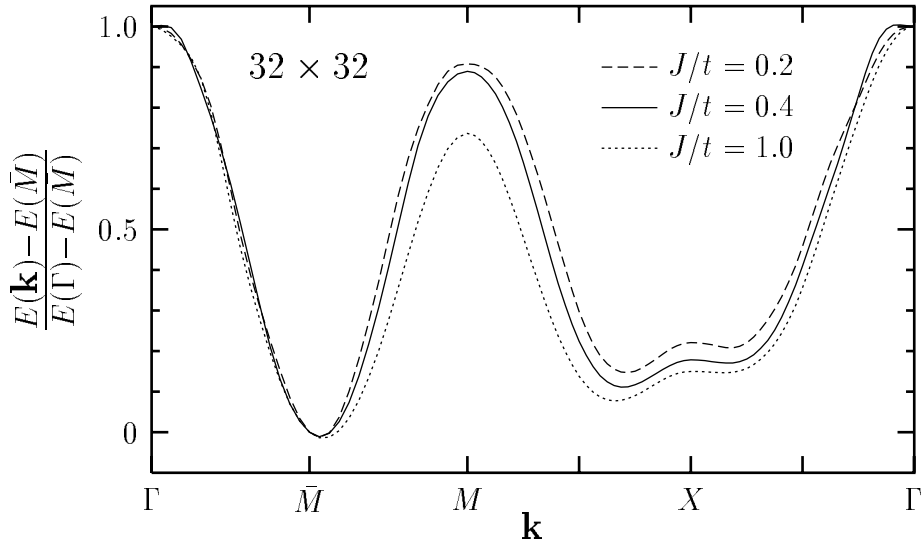
\begin{figure}
  \caption{The quasiparticle band structure
		at representative values of $J/t$.
	The result was obtained using a $32 \times 32$ lattice.
	The actual scales are as follows.
	For $J/t = 0.2$, $E(\Gamma) - E(\bar{M}) = 0.45t$;
	for $J/t = 0.4$, $E(\Gamma) - E(\bar{M}) = 0.83t$;
	for $J/t = 1$,   $E(\Gamma) - E(\bar{M}) = 1.40t$.
	See also Fig.~\protect\ref{fig:band3d}.
}
  \label{fig:band32}
\end{figure}

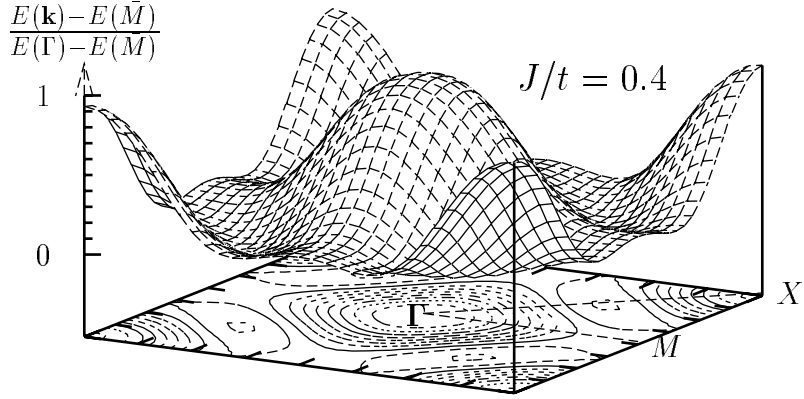
\begin{figure}
  \caption{A three-dimensional plot of the ``normalized'' dispersion
		at $J/t = 0.4$.}
  \label{fig:band3d}
\end{figure}

The dispersion relation calculated on a $32 \times 32$ lattice
  is shown in Figs.\ \ref{fig:band32} and~\ref{fig:band3d}.
Of course,
  the most notable difference from previous slave-fermion studies
  is that
  there is no degeneracy between $\bk$ and $\bk + \bQ$.
In the low-energy part of the band
  there is now
  an extended nearly flat region near $X = (\pi, 0)$.
This is in agreement
  with the experimental
  angle-resolved photoemission spectroscopy (ARPES) data
  and is probably key
  to explaining some
  of the experimental properties of the cuprates.\cite{DNB94}

Another effect observed in the results
  is that 
  the band minimum
  slightly shifts away from
  $\bar{M} = (\pi/2, \pi/2)$ towards $M = (\pi, \pi)$
  while the maximum
  splits,
  moving from $\Gamma = (0, 0)$ in the direction of $X = (\pi, 0)$ and $Y$.
Using various lattices
  and supplementing it with interpolation by means
  of trigonometric-functions fits,
  I found the minimum to be
  at $(0.503\pi, 0.503\pi)$ for $J/t = 0.4$
  and
  at $(0.545\pi, 0.545\pi)$ for $J/t = 1$.

Higher-order terms
  will probably lead to a renormalization
  of the values of $t$ and $J$,
  similar to what was found
  in the slave-fermion case.\cite{liu-manousakis}
In the present study, the shape of the dispersion
  is almost the same in a fairly wide parameter range
  and is presumably very accurate
  but the absolute numbers may change.

\subsection{The two-hole problem}

As a further test of the method's usefulness,
  I applied it to the two-hole problem.
Many issues that come up here
  are the same as in the many-body problem
  and, again, availability of some numerical data
  makes it
  a good opportunity to test viability and/or accuracy
  of the diagrammatic approach
  in a context more complex than the single-particle one.
To the best of the author's knowledge,
  no such test has been made
  in the framework of the slave-fermion approach either
  and I believe this problem
  must be definitively solved analytically
  before attacking the many-body problem.
All the following consideration
  is restricted to pairs with total momentum zero.

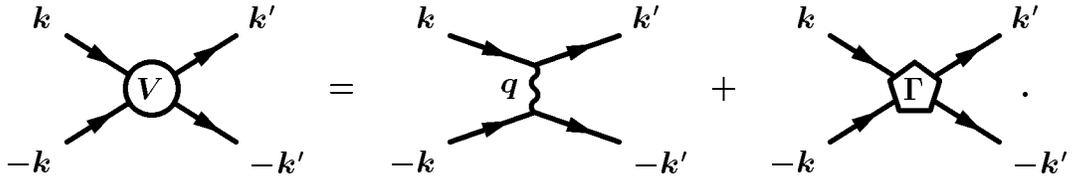
\begin{figure}
  \caption{The diagrams included in the four-particle scattering amplitude
		$V(\bk, \bkp, \omega)$.
	The next-order omitted diagrams are shown in the following
		Fig.~\protect\ref{fig:omitted}.
  }
  \label{fig:vertex}
\end{figure}

The bound state
  is found by solving the Bethe-Salpeter equation
\begin{equation}
\label{eq:bethe-salpeter}
  G_{\uparrow\downarrow} (\bk, \omega)
	= \frac{1}{2\pi N}
	 \sum_\bkp \int \! d \omega' \,
		V (\bk, \bkp, \omega - \omega')\,
		G(\bkp, E + \omega')\, G(-\bkp, E - \omega')\,
		G_{\uparrow\downarrow} (\bkp, \omega'),
\end{equation}
  where $E$ is half the total energy
  ({\it i.e.\/} the energy {\em per hole\/}).
The function $G_{\uparrow\downarrow} (\bk, \omega)$
  appears from decoupling the four-particle hole Green's function
  at the bound state.
It corresponds to the anomalous Green's function
  in the many-body (Eliashberg) problem
  and to the pair wave function (in momentum space)
  in the limit of static hole-hole interaction.
For the four-particle scattering amplitude
  I used
\begin{displaymath}
  V(\bk, \bkp, \omega)
	 = (zt)^2 M (\bk, \bk - \bkp)\, M (\bkp, \bkp - \bk) \,
		 D(\bk - \bkp, \omega)
	   + \Gamma_{\text{cont}} (\bk - \bkp),
\end{displaymath}
  where
  $D(\bq, \omega) = 2\omega_\bq (\omega^2 - \omega_\bq^2 + i0)^{-1}$
  is the magnon Green's function.
Note that the product of $M$'s is simply
\begin{math}
    \left[ \gamma_\bk \gamma_\bkp
    - \case{1}{2} \gamma_\bq (\gamma_\bk^2 + \gamma_\bkp^2)\right]
		 \nu_\bq^{-1}.
\end{math}
This amplitude
  corresponds to the leading-order spin-wave exchange diagram
  and the contact interaction,
  which is all
  diagrammatically represented in Fig.~\ref{fig:vertex}.

In solving Eq.~(\ref{eq:bethe-salpeter}),
  the convolution over frequencies
  was performed using fast Fourier transforms (FFT),
  taking in various regimes
  1024 to 8192 points for the frequency mesh.
The frequency cutoff
  was typically $8t + 3J$.
Although despite appearances
  the structure of the vertex allows FFT over momenta as well,
  doing so in practice involves fairly large overheads
  and small lattices
  are solved faster by direct summation in $\bk$ space,
  using all available symmetries.
The bound state energy is found as
  the $E$ at which
  (the real part of) the largest eigenvalue for a given symmetry
  becomes equal to 1
  (the imaginary part was kept under $0.05$
   and its negligible influence was verified).
The binding energy per pair
  is then $\Delta_B = 2(E - E_{\text{min}})$,
  where $E_{\text{min}}$
  is the minimum single-hole energy
  (for consistency,
   this must be found
   from the momenta actually available on the cluster
   instead of by interpolation
   or by using the bulk limit result).

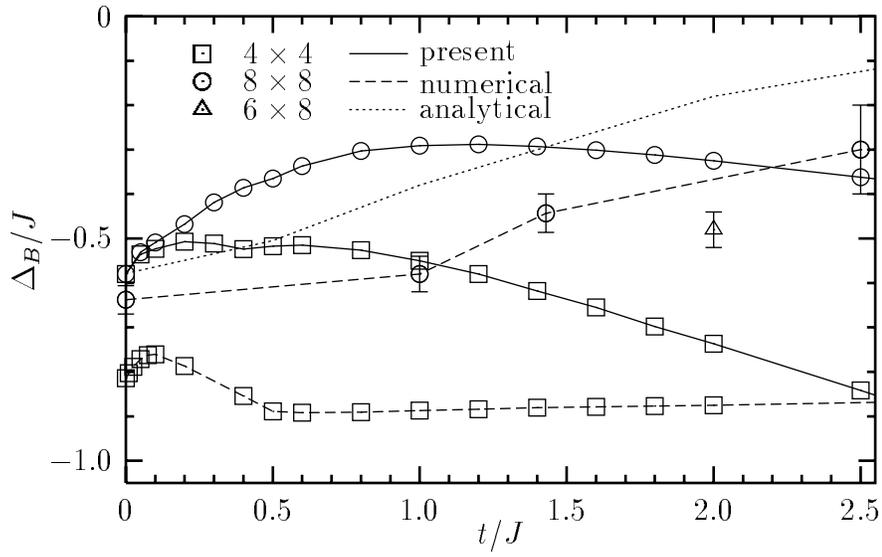
\begin{figure}
  \caption{The two-hole binding energy $\Delta_B$ as a function of $t/J$.
	The numerical results on the $4 \times 4$ lattice
		are from exact diagonalizations.
	The numerical results on the $8 \times 8$ lattice
		are from Monte Carlo studies
		(Barnes and Kovarik\protect\cite{barnes-kovarik90}
		 for $t/J = 0$
		 and
		Boninsegni and Manousakis\protect\cite{boninsegni-manousakis93}
		 for the other points).
	The numerical result on the $6 \times 8$ lattice
		was quoted by
		White and Scalapino\protect\cite{white-scalapino96}
		in a density matrix renormalization group study.
	The analytical result is from Ref.~\protect\onlinecite{BCS96}.
	The irregularities at small $t/J$ on the $4 \times 4$ lattice
		are caused by level crossings
		(at $t/J = 0.4814$ and $0.1526$, see
		Ref.~\protect\onlinecite{barnes-etal92}).
   }
  \label{fig:binding}
\end{figure}

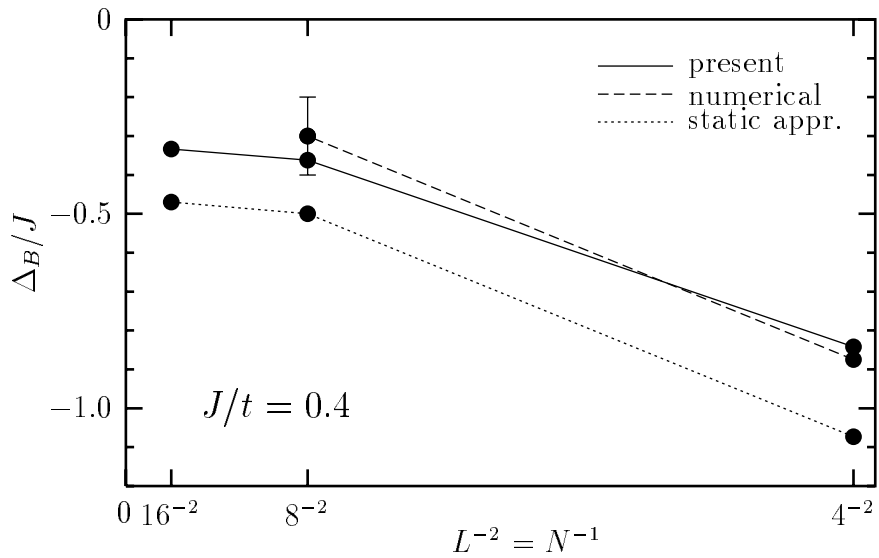
\begin{figure}
  \caption{Size dependence of the two-hole binding energy $\Delta_B$
		at $J/t = 0.4$.
		$L \times L$ is the lattice size.
	The dotted line
	   is the result obtained using
	   the static approximation described in the text.}
  \label{fig:sizedep}
\end{figure}

The solution I consider is even in frequency
  and has a $d_{x^2 - y^2}$-wave spatial symmetry.
The results, presented in Fig.~\ref{fig:binding},
  show a maybe-satisfactory agreement with numerical data.
The dependence of binding energy on $J/t$
  is noticeably different.
The size dependence
  is however matched rather well,
  suggesting that the difference
  is due to a ``renormalization.''
This is further illustrated in Fig.~\ref{fig:sizedep},
  where the result for obtained on the $16 \times 16$ lattice
  is also plotted,
  demonstrating almost complete finite-size convergence
  at such sizes.

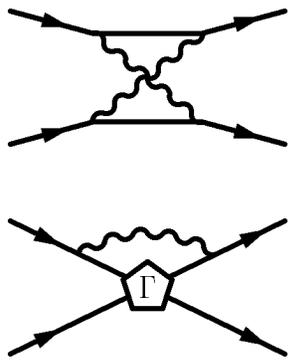
\begin{figure}
  \caption{The higher-order diagrams
		for the four-particle scattering amplitude.
	Both diagrams are expected to be suppressed (see the text).}
  \label{fig:omitted}
\end{figure}

Barring the possibility of an error,
  the most likely cause of the discrepancy
  is contribution of higher-order diagrams.\cite{manySpinWaves}
In the regime of small $t/J$, the calculated binding energy
  behaves as square of the quasiparticle residue,
  which is what is expected but which is not what
  is seen in numerical data.
The small size of the bound state
  may mean that a real-space based approach
  may be more efficient
  as the complexity of diagrams grows very quickly.  
The next-order omitted diagrams
  are shown in Fig.~\ref{fig:omitted}.
The first crossing diagram, however,
  is expected to be zero
  for the same kinematic reason
  as in the single-hole case.\cite{oleg}

Note also that the binding energy
  is quite small relative to the total energy scale.
The result
  is very sensitive to values of interaction
  and the values of single-hole energy.
Given the complexity of the problem,
  renormalizing the interaction
  is an attractive option.

Various features of the two-hole bound state
  have been discussed at length in the literature
  (see Ref.~\onlinecite{dagotto94} and references therein,
  recent references
  are Refs.\ \onlinecite{white-scalapino96},
  \onlinecite{BCS96},
  and the very detailed Ref.~\onlinecite{riera-dagotto97})
  and will not be described here
  except for a note on the ``static'' nature of the bound state.
In the most naive, ``nonrelativistic'' static limit
  the spin-wave Green's function
  is replaced by $ - 2 \omega_\bq^{-1}$,
  thus creating a potential interaction.
A more accurate approximation
  is to take into account holes' velocity
  but assume a static Green's function,
\begin{math}
  G_{\uparrow\downarrow} (\bk, \omega)
   = G_{\uparrow\downarrow} (\bk, \omega = 0)
   \equiv \Psi (\bk).
\end{math}
This approach,
  used in several studies of
  the \tj model,\cite{oldStuff}
  is based on the following argument.
Since the vertex function $M (\bk, \bq)$
  is small for small $q$,
  magnons with small momentum are not important in the problem.
The {\em typical\/} energy of magnons
  involved in spin-wave exchange
  is of the order of $2J$.
On the other hand, 
  the energy scale for holes excitations
  is $\case{1}{2} (|\Delta_B| + W_{\text{eff}}) $,
  where the effective hole bandwidth
  is $W_{\text{eff}} \sim 0.3t$,
  so that even at $J/t \sim 0.4$
  the holes may be regarded as very slow.
Mathematically it means that
  in the range in which $G (\bk, E + \omega) G(-\bk, E - \omega)$
  is not negligible,
  the function $G_{\uparrow\downarrow} (\bk, \omega)$
  does not change much,
  so that for integration in Eq.~(\ref{eq:bethe-salpeter})
  one can replace $G_{\uparrow\downarrow} (\bk, \omega)$
  with $G_{\uparrow\downarrow} (\bk, 0)$.
The results of solving
  the static version of the Bether-Salpeter equation,
\begin{equation}
\label{eq:bethe-salpeter-static}
  \Psi (\bk)
	= \frac{1}{2\pi N}
		\sum_\bkp \int\! d \omega'\,
		V (\bk, \bkp, -\omega') \,
		G(\bkp, E + \omega')\,G(-\bkp, E - \omega')\,
		\Psi (\bkp),
\end{equation}
  at $J/t = 0.4$ are shown in Fig.~\ref{fig:sizedep}.
They show that the static approximation,
  of course far from being completely safe,
  should be sufficient for most estimations,
  especially considering that
  errors introduced this way
  seem to be less than
  those originating from other sources
  (presumably from
   neglecting or mean-field-decoupling higher-order diagrams).

\section{Summary} \label{sec:summary}

I have analysed the \tj model
  using an ``unconstrained'' representation,
  in which fermions and bosons are fully commuting
  and
  in which fermions correspond closer
  to the fermions in the original \tj model.
The results for single-hole properties
  within noncrossing-diagram approximation
  have been found to be in good agreement with numerical data.
The results clarify the nature of the hole dynamics.
It is almost completely described by only two kinds
  of underlying processes:
  $(i)$ a string-picture-like motion by means of emitting/absorbing
  a spin wave at each step
  and $(ii)$ direct hopping on ground-state spin fluctuations.
The results for the two-hole problem
  are less clear and more work may be required.

In the physical regime of small $J/t$ and on antiferromagnetic background,
  the numerical differences
  of the slave-fermion formulation,
  from the presented method are fairly small.
However since the described method
  has exactly the same level of complexity
  (or rather simplicity),
  there is no reason not to use it.
It is also important that this representation
  has the potential to be used on a background
  with an arbitrary magnetic order.


\acknowledgments

The author's work
  was supported from an Australian Research Council grant.
Part of the work
  was performed
  at das Max-Planck-Institut f\"ur Physik komplexer Systeme
  in Dresden, Germany.


\appendix
\section{The motion of a hole\protect\\
	on an Ising-type background}
\label{sec:appendix}

In this Appendix
  I describe the hole motion
  in the presence of ground-state spin fluctuations
  using a very simple one-dimensional Ising system.
In the following,
  all spin-down-sublattice sites have been rotated by $\pi$.

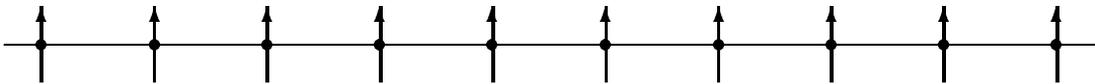
\begin{figure}
  \caption{The perfectly antiferromagnetic ground state.
		In this and following figures half of the spins have been
		rotated by $\pi$.}
  \label{fig:isingGround}
\end{figure}

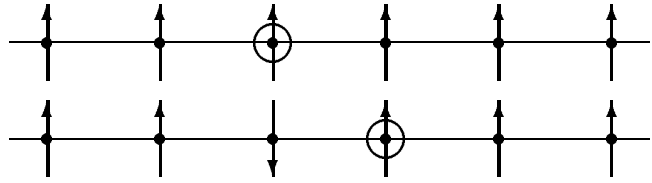
\begin{figure}
  \caption{Motion of a holon in perfectly antiferromagnetic background.}
  \label{fig:isingMotion}
\end{figure}

The perfect antiferromagnetic ground state
  is shown in Fig.~\ref{fig:isingGround},
  and the motion of a hole in such a background
  is depicted in Fig.~\ref{fig:isingMotion}.
As the hole moves,
  it leaves behind a trail of flipped spins
  (in one dimension they all
   amount to one spin-order distortion at the origin).
The hole
  may move back to wipe out the distortion it created.
In the corresponding two-dimensional, string picture,
  the hole
  creates a spin distortion at {\em each\/} step away from the origin.
The hole
  moves by either emitting a spin wave
  or by absorbing one that it has created earlier.
The spin wave
  can be thought of as a complex-spread-out-in-space
  form of spin flip/distortion,
  conceptually the two are the same.
Notice that the ghost spin field on the hole site
  is left effectively idle,
  always being nonflipped according to the convention.

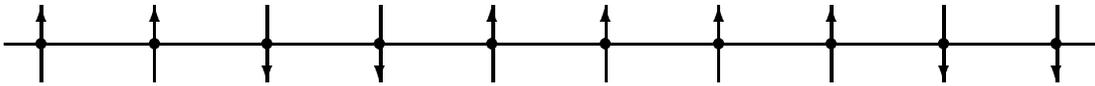
\begin{figure}
  \caption{An antiferromagnetic ground state with spin fluctuations present.}
  \label{fig:neelGround}
\end{figure}

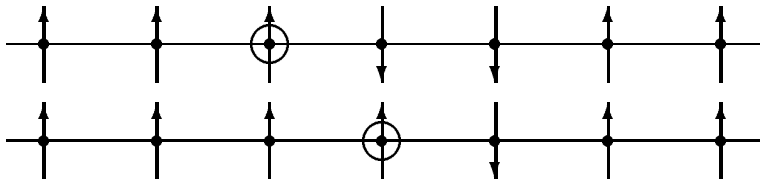
\begin{figure}
  \caption{Motion of a holon encountering a ground-state spin fluctuation.}
  \label{fig:neelMotion}
\end{figure}

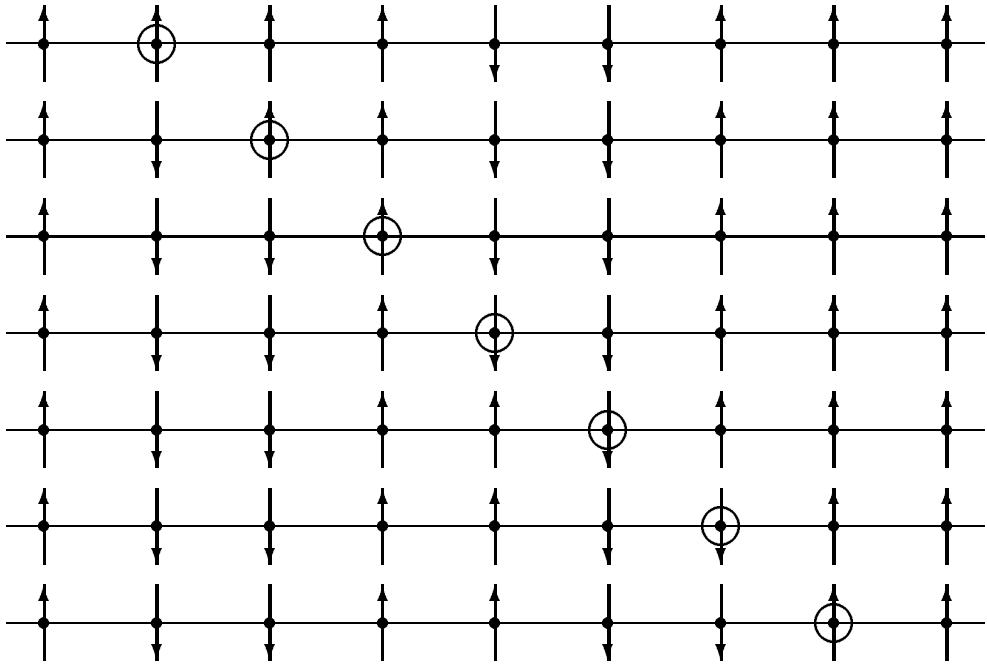
\begin{figure}
  \caption{Motion of a hole in a background with spin fluctuations,
		as appearing in the unconstrained representation.}
  \label{fig:theBigOne}
\end{figure}

Now assume that for whatever reason
  (perhaps due to a kind of frustration)
  the ground state
  has the form in Fig.~\ref{fig:neelGround},
  that is it has spin flips/fluctuations frozen in it
  (incidentally, the flips come in pairs).
On this background, in addition to the processes
  shown earlier in Fig.~\ref{fig:isingMotion},
  there will be occasional processes
  where the hole moves without creating any new spin distortions,
  but only shifting those already present (Fig.~\ref{fig:neelMotion}).
Analyzing the process in Fig.~\ref{fig:neelMotion},
  an inconvenience of the holon approach becomes obvious.
Namely, it is that
  although no actual spin excitation was created,
  there is a spin flip in the spin field
  due to rigidity of the ghost spin attached to the holon.
This makes the analytical transformation unnatural
  and it will be hard to account for such processes accurately.
The easy way out
  is to fully employ the ghost spin by forcing
  it flip when necessary,
  thus carrying much more information and doing much more work.
Figure~\ref{fig:theBigOne} illustrates
  the motion of a hole as described by the new conventions.
The direction of the ghost spin
  effectively indicates
  whether any encountered spin flip could have been
  created by the hole back in time
  or if it is a background spin fluctuation.
In the latter case,
  the hole advances to the next site free---no strings attached.



\begin{figure}
  \caption{The Hilbert space in the \tj model.
	   The arrows represent electrons.}
  \label{fig:hilbert}
\end{figure}

\begin{figure}
  \caption{The Hilbert space in the \tj model as represented
		in the slave-fermion formulation.
	   The arrows represent the spin field $\bspin_n$,
		while the circles show holons.
	   Note that the last two configurations are on certain sublattices.}
  \label{fig:holonHilbert}
\end{figure}

\begin{figure}
  \caption{Graphical representation of the hopping part
		of the Hamiltonian
		in the unconstrained model
		(the sublattice-unsymmetric version).}
  \label{fig:processes}
\end{figure}

\begin{thebibliography}{10}
\bibitem{manySpinWaves}
Another potential reason for discrepancy
  is more subtle.
Consider for example the case of $J/t \to 0$.
In solving the problem analytically,
  it is easy to generate a very large number of virtual spin waves,
  while in fact physically it is clear
  that the number has to be less than the number of lattice sites.
The diagrammatic approach
  is inherently for the bulk limit,
  which means that a direct comparison is not always possible.
The actual criterion for agreement should be then that
  when no finite size effects are found in numerical data,
  then the results must agree.
Otherwise the procedure is undefined,
  and some conclusions can only be drawn indirectly.
\end{thebibliography}
\end{document}